\documentclass[prb,superscriptaddress,floatfix,twocolumn]{revtex4}
\usepackage{epsfig}

\begin{document}

\title{Raman Scattering versus Infrared Conductivity: Evidence for 1D Conduction
in La$_{2-x}$Sr$_{x}$CuO$_{4}$}

\author{F. Venturini}
\affiliation{Walther Meissner Institut, Bayerische Akademie der
Wissenschaften, 85748 Garching, Germany}
\author{Q.-M. Zhang\cite{qm}}
\affiliation{Walther Meissner Institut, Bayerische Akademie der
Wissenschaften, 85748 Garching, Germany}
\author{R. Hackl}
\affiliation{Walther Meissner Institut, Bayerische Akademie der
Wissenschaften, 85748 Garching, Germany}
\author{A. Lucarelli}
\author{S. Lupi}
\author{M. Ortolani}
\author{P. Calvani}
\affiliation{INFM and Dipartimento di Fisica - Universit\`a di Roma
``La Sapienza'', Piazzale Aldo Moro 2, 00185 Roma, Italy}
\author{N. Kikugawa}
\author{T. Fujita}
\affiliation{ADSM, Hiroshima University, Higashi-Hiroshima 739-8526, Japan}
\date{\today}

\begin{abstract}
Raman and Infrared (IR) spectra of an underdoped
La$_{1.90}$Sr$_{0.10}$CuO$_{4}$ single  crystal have been measured
as a function of temperature. Both techniques provide
unconventional low-energy spectra. The IR conductivity exhibits
features peaked at finite frequencies which do not have a
counterpart in the Raman response. Below approximately 100~K a
transfer of both Raman and IR spectral weight towards lower
energies is found and a new component in the Raman response builds
up being characterized by a very long lifetime of electrons
propagating along the Cu-O bonds.
\end{abstract}

\maketitle

In conventional metals infrared (IR) and dc ($\omega=0$)
conductivity as well as electronic Raman scattering are expected
to reveal a consistent picture of the electronic properties even
in lowest order approximation. In metals with strong electronic
interactions the spectral response will be characteristically
renormalized (vertex corrections) for the IR and the Raman
response. In spite of the strong correlations in high-$T_c$
superconductors such as YBa$_2$Cu$_3$O$_{6+y}$ (Y123) or
Bi$_2$Sr$_2$CaCu$_2$O$_{8+y}$ (Bi2212) the electron dynamics
derived from IR and Raman spectroscopies agree quite reasonably as
long as the selection rules are properly taken into
account.\cite{opel2000} On the other hand, in low-$T_c$ cuprates
such as La$_{2-x}$Sr$_{x}$CuO$_{4+y}$ (LSCO) the far-infrared
conductivity $\sigma (\omega, T)$ exhibits strong additional
contributions\cite{Lobo,basov} peaked between 100 and 500~$\rm
cm^{-1}$ which do not have a counterpart in the Raman
spectra.\cite{naeini}  In Bi$_{2}$Sr$_{2}$CuO$_6$ (Bi2201) a
similar peak in $\sigma (\omega, T)$ has been observed at low
temperatures which is well separated from a conventional Drude
term\cite{lupi00} having a width of only a few $\rm cm^{-1}$. This
indicates the existence of states with a very long lifetime well
below the pseudogap temperature \cite{Timusk} $T^{\ast}$ and above
the superconducting transition at $T_c$. So far there is neither
an explanation for this type of behavior, nor experimental
information from a complementary spectroscopy.

In  the present paper we compare IR and Raman spectra taken on the
very same surface of a high-quality single crystal of underdoped
LSCO. We observe strong absorption features in the far-infrared
and, in the Raman response, a novel component for a given
combination of photon polarizations. We show that these IR and
Raman anomalies are related with each other and consistently point
towards one-dimensional conductivity in the Cu-O planes.

The LSCO single crystal was grown by the traveling-solvent
floating-zone  technique. The Sr content was determined with
energy-dispersive X-ray spectroscopy to be $0.10 \pm 0.01$. The
resistively measured transition temperature is $T_c = 28$~K with a
width (10-90\%) $\Delta T_c = 1$~K. After polishing the sample was
reannealed in flowing oxygen for 50 hours at 920$^{\circ}$C,
slowly (50 hours) cooled down to 500$^{\circ}$C where it was kept
for another 50 hours. Finally it was cooled down to ambient
temperature (50 hours). This procedure ensured that the crystal
surface was strain free and of comparable quality as those used
for photoemission experiments.

In the infrared experiment, the sample was mounted on the cold
finger of a two stage closed-cycle cryostat, whose temperature was
kept constant within $\pm 2$~K and could be varied from 20 to
295~K. The reflectance $R(\omega,T)$ of the crystal was measured
at quasi-normal incidence (8$^{\circ}$) with the radiation field
polarized in the CuO$_2$ ($a-b$) planes. The reference was
obtained by evaporating  a gold layer onto the sample by using a
hot filament placed in front of the surface. \cite{Homes} Spectra
from 20 to 20,000~cm$^{-1}$ were collected by a rapid scanning
interferometer.  In the far infrared, different combinations of
mylar beamsplitters and bolometers were used to exclude that
``ghost'' spectral features  might  appear when connecting
different spectral ranges. For the Raman experiment the sample was
mounted on the cold finger of a He-flow cryostat. The spectra were
measured with a resolution of $10~{\rm cm^{-1}}$. All
polarizations were in the $a-b$ plane. For different polarization
combinations of the incoming and the outgoing photons different
regions on the Fermi surface can be probed independently.

\begin{figure}
{\hbox{\centerline{\psfig{figure=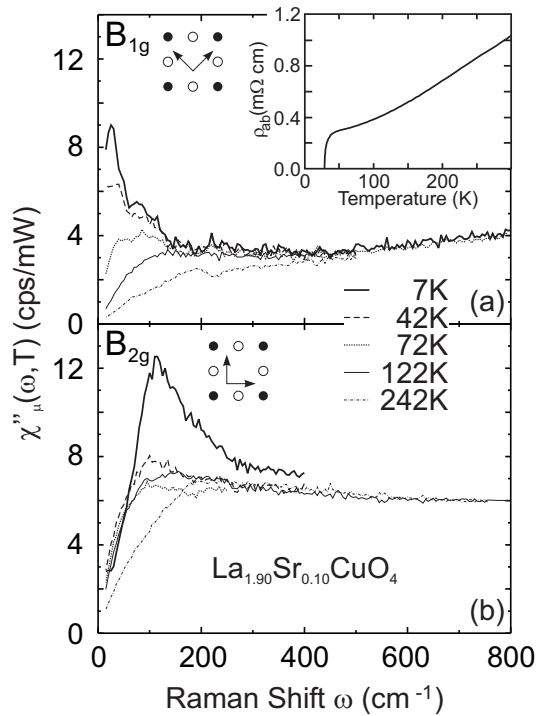,width=7.0cm}}}}
\caption[]{Raman response $\chi_{\mu}^{\prime
\prime}(\omega,T)$ of La$_{1.90}$Sr$_{0.10}$CuO$_4$. Spectra for
the $B_{1g}$ a) and the $B_{2g}$ b) configuration are shown for
the normal and the superconducting state at the indicated
temperatures. The inset in a) shows the resistivity of the sample
between  4.2 and 300~K.} \label{fig-raman}
\end{figure}

Spectra of the normal-state electronic Raman response
$\chi_{\mu}^{\prime \prime}(\omega, T)$ of
La$_{1.9}$Sr$_{0.1}$CuO$_{4}$ are shown in Fig.~\ref{fig-raman}
for the two symmetries $\mu = B_{1g},B_{2g}$ projecting out
electrons with momenta approximately along the principal axes and
the diagonals of the CuO$_2$ plane, respectively. The structures
from vibrational excitations are very weak in the $a-b$~plane
(typically 10\% of the total intensity) and have been subtracted
out. At high energies, $\hbar \omega \geq 400~{\rm cm^{-1}}$, the
spectra are qualitatively similar to earlier results \cite{naeini}
and to those measured in Y123 and Bi2212.
\cite{opel2000,chen93,Sugai} Up to approximately $1500~{\rm
cm^{-1}}$ the continuum is essentially constant in $B_{2g}$
symmetry (Fig.~\ref{fig-raman}~(b)) and increases linearly in
$B_{1g}$ symmetry (Fig.~\ref{fig-raman}~(a)).
In the range below $200~\rm cm^{-1}$ we observe the usual increase
of the spectral weight in $B_{2g}$ symmetry upon reducing the
temperature (Fig.~\ref{fig-raman}~b)) leading to an increasing
slope of the response in the dc limit. Since $\partial
\chi_{\mu}^{\prime \prime}(\omega \rightarrow 0, T)/\partial
\omega \propto \tau(T)$ with $\tau(T)$ the static lifetime of the
quasiparticles \cite{opel2000} this is compatible with the
metallic decrease of the resistivity. In the superconducting state
the characteristic redistribution of spectral weight and the
formation of a pair-breaking peak is found.

In $B_{1g}$ symmetry a similar metallic increase towards low
temperature is revealed between approximately 250 and 100~K. The
spectra are more flat indicating a larger relaxation rate or a
shorter lifetime as well as a weaker temperature dependence than
in $B_{2g}$ symmetry. Similarly as in previous studies no
pair-breaking feature can be observed below $T_c$ at this doping
level. \cite{opel2000,naeini,Sugai} Completely unexpectedly,
however, the low-energy response increases strongly for $T<100$~K.
Although the spectra are measured down to $15~{\rm cm^{-1}}$ the
linear decrease (inevitably present for causality) towards zero
energy cannot be resolved any more. Consequently, the lifetime
must become very long and exceed the one observed at $B_{2g}$
symmetry by far. The reason why this has not been observed before
in LSCO is most likely because the spectra were not measured at
energies below $80-100~{\rm cm^{-1}}$. For Y123 and Bi2212 such
type of an increase can be excluded, hence we are facing a type of
electron dynamics which has escaped observation so far. Are there
corresponding features in the IR conductivity?

\begin{figure}
{\hbox{\centerline{\psfig{figure=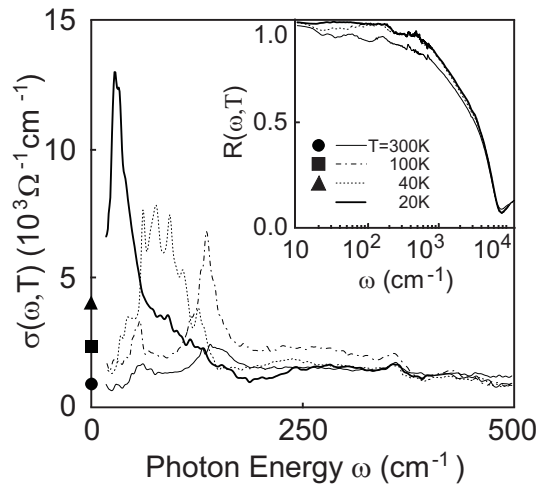,width=7.0cm}}}}
\caption[]{In-plane optical conductivity
$\sigma(\omega,T)$ of La$_{1.90}$Sr$_{0.10}$CuO$_4$. The
underlying reflectivity data are shown in the inset. The symbols
at zero energy represent the dc conductivity values at the
respective temperatures.} \label{fig-ir}
\end{figure}

The raw reflectivity $R(\omega,T)$ of the
La$_{1.90}$Sr$_{0.10}$CuO$_4$ crystal is shown in the inset of
Fig.\ \ref{fig-ir} at different temperatures. The real part of the
optical conductivity $\sigma (\omega,T)$ was extracted from
$R(\omega,T)$ by usual Kramers-Kr\"onig transformations. A
Drude-Lorentz fit was used to extrapolate the reflectivity to
$\omega = 0$. On the high-energy side, the data were extrapolated
with the $R(\omega)$  reported in Ref. \onlinecite{Tajima}. The
resulting infrared conductivity $\sigma (\omega,T)$ of
La$_{1.90}$Sr$_{0.10}$CuO$_4$ is shown in Fig.\ \ref{fig-ir}. At
all temperatures one observes  strong spectral features below
200~cm$^{-1}$ that are not related to infrared-active lattice
vibrations of LSCO in the $a-b$ plane, as a comparison with the
weak phonon line at 360~cm$^{-1}$ demonstrates immediately. They
cannot be wings of a Drude contribution either, because the dc
conductivity at the respective temperatures is much lower than the
conductivity in the peak at finite energies (see Fig.\
\ref{fig-ir} and its caption). Below 100~K the spectral structures
merge into a huge peak, and the conductivity reaches about
13,000~$\Omega^{-1}$~cm$^{-1}$ at 40~cm$^{-1}$ and 20~K. This
softening goes along with a transfer of spectral weight which
depletes the region between 100 and 200~cm$^{-1}$.
The resulting  peak is similar to those observed in systems with
one-dimensional (1D) charge ordering such as organic conductors
\cite{Schwartz} and ladders\cite{Osafune}.  In the latter
compounds,  an infrared pseudogap opens at low $T$ along the $a$
axis that is parallel to the rungs, while a gapless strong peak is
observed along the $c$ axis, parallel to the legs.\cite{Osafune}.
The latter behavior is remarkably similar to that reported in
Fig.\ \ref{fig-ir}. Strong infared peaks at finite frequencies are
also reported for other superconducting cuprates with low $T_c$
such as La$_2$SrCuO$_{4.06}$,\cite{Lobo}
Nd$_{2-x}$Ce$_{x}$CuO$_{4-y}$,\cite{lupi99}
Bi$_{2}$Sr$_{2}$CuO$_6$, \cite{lupi00}, and assigned to isolated,
interacting, and ordered polarons, respectively. In overdoped
La$_{2-x}$Sr$_x$CuO$_4$ similar far-infrared features were
observed. They were attributed to disorder and analyzed in terms
of an ``anomalous Drude'' model.\cite{Startseva}

In the present work, the comparison of IR and Raman data allows us
to get deeper insight into the origin of the strong absorption
peaks of LSCO shown in Fig.\ \ref{fig-ir}. The fully developed
pair-breaking feature in the $B_{2g}$ Raman spectrum (Fig.\
\ref{fig-raman}~(b)) indicates that the sample is very clean. In
the presence of disorder the spectral weight at approximately
twice the maximal gap energy $2\Delta_0$ would be reduced
\cite{tpd92,tpd95}, and the pair-breaking peaks would be smeared
out rapidly and disappear for $\hbar/\tau \simeq \Delta_0$. First
of all, we can therefore exclude that the high conductivity at
approximately $40~{\rm cm^{-1}}$ (Fig.\ \ref{fig-raman}~(a))
originates from the presence of a significant amount of
impurities. Secondly, the low-energy low-temperature Raman
response in $B_{1g}$ symmetry shows a pronounced increase of the
lifetime of quasiparticles propagating along the principal axes
(Fig.\ \ref{fig-raman}~(a)). This high ``Raman conductivity'' in
the dc limit goes along with the occurrence of the conductivity
peak in the IR at low but finite energies (Fig.\ \ref{fig-ir}).

\begin{figure}
{\hbox{\centerline{\psfig{figure=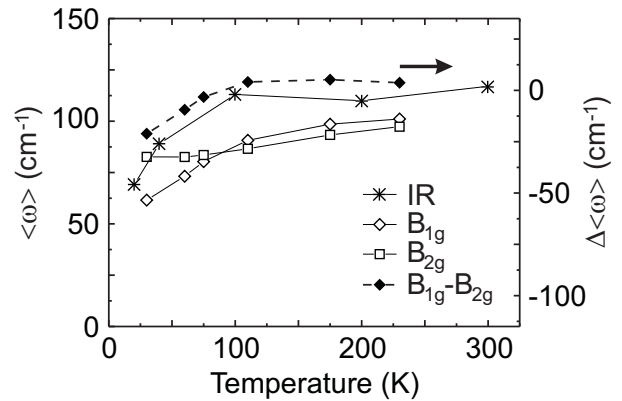,width=8cm}}}}
\caption[]{First moments (left scale) of the IR
(asterisks) and Raman (open symbols) spectra. At temperatures $T
\leq 100$~K spectral weight is transferred to low energies in the
IR and in the $B_{1g}$ Raman spectra. For better visualization we
plot the difference between the $B_{1g}$ and the $B_{2g}$ moments
$\Delta \langle \omega \rangle$ (full diamonds, right scale).}
\label{fig-average}
\end{figure}

The relationship between the two phenomena can be quantified by
calculating the spectral weight. The quantity
\begin{equation}
<\omega>_{IR} \,=\, {{\int_{\omega_1}^{\omega_2} \omega \,
\sigma  (\omega)\, d \omega}
\over {\int_{\omega_1}^{\omega_2} \sigma (\omega) \, d \omega}} \, .
\label{moment-ir}
\end{equation}
\noindent is the first moment of the infrared spectrum between
$\omega_1$ and $\omega_2$. In lowest order one obtains
$\chi^{\prime \prime}(\omega, T) \propto \omega \sigma(\omega, T)$
for the Raman response \cite{opel2000,viro92}. Hence, we determine
\begin{equation}
<\omega>_{\mu} \,=\, {{\int_{\omega_1}^{\omega_2} \, d \omega} \,
\chi_{\mu}^{\prime \prime}(\omega, T)  \over
{\int_{\omega_1}^{\omega_2} \, {{d \omega} \over {\omega}} \,
\chi_{\mu}^{\prime \prime}(\omega, T)}}. \label{moment-raman}
\end{equation}
In Fig.\ \ref{fig-average}, all moments are plotted vs.
temperature using ${\omega_1}$ = 20~cm$^{-1}$ and $\omega_2$ =
200~cm$^{-1}$. The moment derived from the conductivity is
temperature independent down to approximately 100~K and decreases
by up 35\% for $T<100$~K. The moments of the Raman response do not
have a constant part since there is no sum rule. At $B_{2g}$
symmetry an essentially linear decrease is found while at $B_{1g}$
symmetry a crossover of two linear regimes with different slopes
occurs. In the high-temperature range the slopes of $B_{1g}$ and
$B_{2g}$ symmetry are similar. In order to better visualize the
changes in $B_{1g}$ symmetry we also plot the difference
$\Delta<\omega> = <\omega>_{B_{1g}} - <\omega>_{B_{2g}}$ in Fig.\
\ref{fig-average}. Apparently, $\Delta <\omega>$ closely follows
the behavior of $<\omega>_{IR}$. Both quantities are nearly
constant at high $T$ and exhibit a kink around 100~K. Of course,
the relative changes cannot be expected to be equal since
different quantities are being measured. However, the comparison
demonstrates that the enhanced conductivity at finite energy
observed in the  IR spectra and the exceptionally long
quasiparticle lifetime derived from the Raman response are
apparently two aspects of a common underlying change in the
electronic properties.
For the similarity of the IR response in LSCO and 1D ladder
compounds\cite{Osafune} we consider charge ordering as a possible
scenario for an explanation of our results. The selection rules
and the coherence factors help clarifying the different spectral
shape of the response. In crystals or subunits such as the CuO$_2$
planes with a center of inversion the mutual exclusion principle
holds, and IR-active (odd or polar) excitations cannot be observed
in the Raman spectrum (even excitations) and vice versa
\cite{tinkham}. In addition, due to the coherence factors
collective phenomena such as superconductivity or density waves
which are a possible way to look at charge ordering lead to
different spectra in the Raman response and in the conductivity
\cite{einzel,dora2000}. For instance, in the superconducting state
the Raman response is characterized by a pile up of spectral
weight around $2\Delta_0$ while the conductivity is partially or
fully suppressed below twice the gap, and the weight is
transferred towards $\omega=0$. In case of a density wave the
roles of IR and Raman response are essentially reversed. For a
pinned density wave there is an additional collective mode
developing inside the gap indicating that a finite energy is
required to trigger an in-phase motion of the charges
\cite{dora2000}. Hence, the IR and Raman spectra observed in
underdoped LSCO can, at least on a qualitative level, be
understood in terms of density wave with incipient pinning.
Recently, in a numerical study on the basis of the $t-J$ model
using parameters realistic for LSCO the strong Drude weight for
the conductivity parallel to the stripes and an incoherent
contribution at finite energy were predicted.\cite{tohyama99}

There is indeed a wealth of experimental evidence for fluctuating
stripes in LSCO \cite{bianconi} and static ones in Nd-doped
LSCO\cite{Tranquada,noda99}. In underdoped LSCO with $x \simeq
0.1$ they are oriented parallel to the CuO bonds (principal
axes).\cite{Fujita} This orientation of the stripes would be
compatible with the observed symmetry dependence in Raman
scattering, i.e. the long lifetime seen in the $B_{1g}$ spectra
where, as opposed to the $B_{2g}$ symmetry, the polarizations of
both the incoming and outgoing photons have a finite projection in
the direction of the 1D structures. Hence, the combination of IR
and Raman spectroscopies provides not only additional evidence for
the formation of stripes in LSCO but also clarifies the low-energy
electron dynamics. The onset temperature for the formation of
stripes is identified by several authors \cite{Sharma,Tsuei} with
the temperature $T^{\ast}$ where a pseudogap,\cite{Timusk}
anisotropic in the momerntum space, \cite{Loeser,Williams} opens
in the charge excitation spectrum. $T^{\ast}$ is of the order of
400 K for LSCO with $x \simeq$ 0.10, consistent with the present
IR observations which already show a well-defined excitation peak
at room temperature. The influence of charge-ordering phenomena on
the electron dynamics increases with decreasing temperature.
Significant changes of the physical properties can be expected
when one-dimensional objects becomes longer than the distance
between the stripes and comparable to the electron mean free path.
Indications of an enhanced dc conductivity along the stripe
direction have indeed been found recently by longitudinal
transport \cite{ando2001} close to temperatures at which the
moments start to decrease (Fig.~\ref{fig-average}). Similarly, the
transverse (Hall) conductivity decreases below approximately 80~K
when the stripes are being completely pinned in Nd-doped
LSCO.\cite{noda99} Finally, recent results of inelastic neutron
scattering on La$_{1.93}$Sr$_{0.07}$CuO$_4$ show\cite{Hiraka} a
strong increase in the stripe fluctuations below 100 K for
energies between 2 and 8 meV (16 and 64 cm$^{-1}$, respectively).

In summary, we have studied a single crystal of underdoped LSCO by
both IR an Raman spectroscopy. At high temperatures the Raman
response is similar to that in other cuprates, while the IR
spectrum shows an anomalous absorption peak around 150 cm$^{-1}$.
Below approximately 100~K a new type of electronic state is
observed to develop in the Raman response, while the IR peak
rapidly displaces to lower energies. The high IR conductivity in
the range between 20 and $200~{\rm cm^{-1}}$ and the high dc
conductivity found in the $B_{1g}$ Raman response are interpreted
in terms of a collective mode and an enhanced 1D dc transport,
respectively, in a charge-density-wave (CDW) scenario or, equivalently,
by low-energy stripe fluctuations with a correlation length of
several lattice constants. It appears that this state is
preferably building up in compounds with low $T_c$ possibly due to
structural peculiarities that favor the pinning of the CDW.\\

We would like to express our gratitude to C. Castellani, C.
Di~Castro, P. Giura,  M. Grilli, A. Nucara, and A. Virosztek for
important discussions and useful comments. Financial support by
the DAAD and the CRUI via the Vigoni program 2000/2001 is
gratefully acknowledged. Part of the project has been supported by
the DFG under grant number HA2071/2-1. F.V. and Q.-M.Z. would like
to thank the Gottlieb Daimler - Karl Benz Foundation and the
Alexander von Humboldt Foundation, respectively.\\

{\it Note added in proof:} Along with or briefly after the
submission of the manuscript two papers on the IR properties of
LSCO samples in a similar doping range appeared.
\cite{takenaka2002,dumm2002} There are certain discrepancies
between the 3 papers which cannot be discussed in detail here.
However, they concern mainly the temperature range where a
far-infrared peak at finite frequency is observed. Indeed, in Ref.
\cite{takenaka2002} a far-infrared peak is resolved from the Drude
term at $T \geq 300$~K  while in Ref. \cite{dumm2002}  the peak is
resolved at low temperature, both in Nd-doped (Fig. 1) and Nd-free
(Fig. 3) LSCO. The discrepancies  among the three papers are
possibly related to a different spectral weight of the Drude term
in different samples. In Ref. \cite{dumm2002} the interpretation
of the peak is slightly different from ours, but also in terms of
charge ordering.

\end{document}